\documentclass{article}
\usepackage{amsfonts}
\usepackage{amssymb}

\input{tcilatex}
\begin{document}

\title{ Modal-Hamiltonian interpretation of quantum mechanics and Casimir
operators: the road toward quantum field theory}
\author{Juan Sebasti\'{a}n Ardenghi \\
CONICET-IAFE-Universidad de Buenos Aires \and Mario Castagnino \\
CONICET-IAFE-IFIR-Universidad de Buenos Aires \and Olimpia Lombardi \\
CONICET-Universidad de Buenos Aires}
\maketitle

\begin{abstract}
The general aim of this paper is to extend the Modal-Hamiltonian
interpretation of quantum mechanics to the case of relativistic quantum
mechanics with gauge $U(1)$ fields. In this case we propose that the
actual-valued observables are the Casimir operators of the Poincar\'{e}
group and of the group $U(1)$ of the internal symmetry of the theory.
Moreover, we also show that the magnitudes that acquire actual values in the
relativistic and in the non-relativistic cases are correctly related through
the adequate limit.
\end{abstract}

\section{Introduction}

In spite of the impressive success of quantum theory, its interpretation is
still an open problem. In previous works (\cite{LC}, \cite{CL}) we have
presented the Modal-Hamiltonian Interpretation (MHI) of non-relativistic
quantum mechanics: a realist, non-collapse interpretation, which defines the
preferred context of the system (the set of the actual-valued observables)
in terms of its Hamiltonian. In subsequent works (\cite{ACL}, \cite{Theoria}%
, \cite{SHPMP}), we have shown that the Modal-Hamiltonian interpretative
rule of actual-value ascription can be formulated in a Galilei-invariant
form when expressed in terms of the Casimir operators of the Galilei group.
In this way, the preferred context selected by the MHI turns out to be
Galilei-invariant, a reasonable result from a realist viewpoint.

Although these interpretative conclusions were obtained for non-relativistic
quantum mechanics, the idea of extending the interpretation to quantum field
theory by replacing the relevant symmetry group sounds rather natural. The
general aim of this paper is, precisely, to open the road toward that
extension by beginning with the case of relativistic quantum mechanics with
gauge $U(1)$ fields, whose symmetry groups are the Poincar\'{e} group and
the internal symmetry group $U(1)$. In this context, we will propose an
interpretative rule according to which the preferred context is defined by
the Casimir operators of the Poincar\'{e} group and of the group $U(1)$ of
the internal symmetry. We will argue that this rule leads to physically
reasonable results in the relativistic realm, since the resulting
actual-valued observables can be considered objective magnitudes because
invariant under the relevant groups. However, one should also expect the
adequate relationship between the results obtained in the relativistic case
and those obtained in non-relativistic quantum mechanics.\ In fact, we will
also show that the magnitudes that acquire actual values in the relativistic
and in the non-relativistic cases are correctly related through the adequate
limit.

For this purpose, the paper is organized as follows.\ In Section 2 we will
briefly introduce the central tenets of the MHI, in particular, the
Modal-Hamiltonian actualization rule of actual-value ascription. On the
basis of the main features of the Galilei group and of its central
extension, presented in Section 3, in Section 4 we will show how the
Modal-Hamiltonian actualization rule can be reformulated under a
Galilei-invariant form in terms of the Casimir operators of the group.
Section 5 will be devoted to the study of the Poincar\'{e} group, its
central extension and the two limits leading to the central-extended Galilei
group: the standard non-relativistic limit and the In\"{o}n\"{u}-Wigner
contraction. In Section 6, the limits of the Casimir operators of the
trivially extended Poincar\'{e} group will be obtained in order to show that
those limits lead precisely to the Casimir operators of the central-extended
Galilei group; this result will count in favor of the extrapolation of MHI
to relativistic quantum mechanics. In Section 7 we will focus on the
internal symmetry $U(1)$, in order to propose that the Casimir operator of
the corresponding symmetry group also acquires an actual-value; in
particular, we will show that such a Casimir operator is the charge which,
as a consequence, may be legitimately considered and actual-valued physical
magnitude. Finally, in Section 8 we will draw our conclusions.

\section{The Modal-Hamiltonian Interpretation}

The MHI belongs to the modal family of interpretations of quantum mechanics
(see \cite{libro modal}); as a consequence, it is a realist interpretation
according to which the quantum state describes the possible properties of a
system but not its actual properties. Here we will only recall the
interpretative postulates relevant to our discussion.

The first step is to identify the systems that populate the quantum
ontology. By adopting an algebraic perspective, a quantum system is defined
in the following terms:

\begin{quotation}
\textbf{Systems postulate (SP):} A \textit{quantum system} $\mathcal{S}$ is
represented by a pair $\mathrm{(}\mathcal{O}\mathrm{,}\,H\mathrm{)}$ such
that (i) $\mathcal{O}$ is a space of self-adjoint operators on a Hilbert
space $\mathcal{H}$, representing the observables of the system, (ii) $H\in 
\mathcal{O}$ is the time-independent Hamiltonian of the system $\mathcal{S}$%
, and (iii) if $\rho _{0}\in \mathcal{O}^{\prime }$ (where $\mathcal{O}%
^{\prime }$ is the dual space of $\mathcal{O}$) is the initial state of $%
\mathcal{S}$, it evolves according to the Schr\"{o}dinger equation in its
von Neumann version.
\end{quotation}

Of course, any quantum system can be partitioned in many ways; however, not
any partition will lead to parts which are, in turn, quantum systems (see 
\cite{Harshman-1}, \cite{Harshman-2}). On this basis, a composite system is
defined as

\begin{quotation}
\textbf{Composite systems postulate (CSP):} A quantum system represented by $%
\mathcal{S}\mathrm{:}\;\mathrm{(}\mathcal{O}\,\mathrm{,}\,H\mathrm{)}$, with
initial state $\rho _{0}\in \mathcal{O}^{\prime }$, is \textit{composite}
when it can be partitioned into two quantum systems $\mathcal{S}^{1}\mathrm{:%
}\;\mathrm{(}\mathcal{O}^{1}\mathrm{,}\,H^{1}\mathrm{)}$ and $\mathcal{S}^{2}%
\mathrm{:}\;\mathrm{(}\mathcal{O}^{2}\,\mathrm{,}\,H^{2}\mathrm{)}$ such
that (i) $\mathcal{O}=\mathcal{O}^{1}\otimes \mathcal{O}^{2}$, and (ii) $%
H=H^{1}\otimes I^{2}+I^{1}\otimes H^{2}$, (where $I^{1}$ and $I^{2}$ are the
identity operators in the corresponding tensor product spaces). In this
case, the initial states of $\mathcal{S}^{1}$ and $\mathcal{S}^{2}$ are
obtained as the partial traces $\rho _{0}^{1}=Tr_{\mathrm{(}2\mathrm{)}}\rho
_{0}$ and $\rho _{0}^{2}=Tr_{\mathrm{(}1\mathrm{)}}\rho _{0}$; we say that $%
\mathcal{S}^{1}$ and $\mathcal{S}^{2}$ are subsystems of the composite
system, $\mathcal{S}=\mathcal{S}^{1}\cup \mathcal{S}^{2}$. If the system is
not composite, it is \textit{elemental}.
\end{quotation}

Since the contextuality of quantum mechanics, implied by the Kochen-Specker
theorem (\cite{K-S}), prevents us from consistently assigning actual values
to all the observables of a quantum system in a given state, the second step
is to identify the preferred context, that is, the set of the actual-valued
observables of the system. For this purpose, we formulate a rule of
actual-value assignment:

\begin{quotation}
\textbf{Actualization rule (AR):} Given an elemental quantum system
represented by $\mathcal{S}\mathrm{:}\;\mathrm{(}\mathcal{O}\,\mathrm{,}\,H%
\mathrm{)}$\textrm{,} the actual-valued observables of $\mathcal{S}$ are $H$
and all the observables commuting with $H$ and having, at least, the same
symmetries as $H$.
\end{quotation}

This preferred context where actualization occurs is independent of time,
since it depends on the Hamiltonian: the actual-valued observables always
commute with the Hamiltonian and, therefore, they are constants of motion of
the system. In other words, the observables that receive actual values are
the same during all the \textquotedblleft life\textquotedblright\ of the
quantum system as such $-$precisely, as a closed system$-$: there is no need
of accounting for the dynamics of the actual properties of the quantum
system as in other modal interpretations (see \cite{Vermaas}).

The fact that the Hamiltonian always belongs to the preferred context agrees
with the many physical cases where the energy has definite value. The
Modal-Hamiltonian actualization rule has been applied to several well-known
physical situations (hydrogen atom, Zeeman effect, fine structure, etc.),
leading to results consistent with experimental evidence (see \cite{LC},
Section 5). Moreover, it has proved to be effective for solving the
measurement problem, both in its ideal and its non-ideal versions (see \cite%
{LC}, Section 6).

\section{The Galilei Group in quantum mechanics}

The space-time symmetry group of non-relativistic $-$classical or quantum$-$
mechanics is the Galilei group $\mathcal{G}$, a Lie group with its
associated Galilei algebra of generators $\mathfrak{g}$. This algebra is
defined by ten symmetry generators $G_{\alpha }$, with $\alpha =1$ to $10$:
one time-displacement $G_{\tau }$, three space-displacements $G_{r_{i}}$,
three space-rotations $G_{\theta _{i}}$, and three Galilei-boost-velocity
components $G_{u_{i}}$, with $i=x,y,z$. The group is defined by the
commutation relations between its generators,%
\begin{equation}
\begin{tabular}{llll}
$\lbrack G_{r_{i}},G_{r_{j}}]=0$ & $(1_{a})$ \ \ \ \ \ \ \ \ \ \ \  & $%
[G_{u_{i}},G_{rj}]=0$ & $(1_{f})$ \\ 
$\lbrack G_{u_{i}},G_{u_{j}}]=0$ & $(1_{b})$ & $[G_{r_{i}},G_{\tau }]=0$ & $%
(1_{g})$ \\ 
$\lbrack G_{\theta _{i}},G_{\theta _{j}}]=i\varepsilon _{ijk}G^{\theta _{k}}$
& $(1_{c})$ & $[G_{\theta _{i}},G_{\tau }]=0$ & $(1_{h})$ \\ 
$\lbrack G_{\theta _{i}},G_{r_{j}}]=i\varepsilon _{ijk}G^{r_{k}}$ & $(1_{d})$
& $[G_{u_{i}},G_{\tau }]=iG^{r_{i}}$ & $(1_{i})$ \\ 
$\lbrack G_{\theta _{i}},G_{u_{j}}]=i\varepsilon _{ijk}G^{u_{k}}$ & $(1_{e})$
&  & 
\end{tabular}
\label{3.1}
\end{equation}%
where $i,j,k=x,y,z$, and $\varepsilon _{ijk}$ is the Levi-Civita tensor,
such that $i\neq k$, $j\neq k$, $\varepsilon _{xyz}=\varepsilon
_{yzx}=\varepsilon _{zyx}=1$, $\varepsilon _{xzy}=\varepsilon
_{yxz}=\varepsilon _{zyx}=-1$, and $\varepsilon _{ijk}=0$ if $i=j$.

Each Galilei transformation $\mathcal{T}_{\alpha }\in \mathcal{G}$ acts on
observables and states as%
\begin{equation}
O\rightarrow O^{\prime }=U_{s_{\alpha }}O\,U_{s_{\alpha }}^{-1}\qquad \qquad
|\varphi \rangle \rightarrow |\varphi ^{\prime }\rangle =U_{s_{\alpha
}}|\varphi \rangle  \label{3.4}
\end{equation}%
where $s_{\alpha }$ is the parameter corresponding to the transformation $%
\mathcal{T}_{\alpha }$, and $U_{s_{\alpha }}$ is the family of unitary
operators describing $\mathcal{T}_{\alpha }$. Since in any case $s_{\alpha }$
is a continuous parameter, each $U_{s_{\alpha }}$ can be expressed in terms
of the corresponding symmetry generator $G_{\alpha }$ as%
\begin{equation}
U_{s_{\alpha }}=e^{iG_{\alpha }s_{\alpha }}  \label{3.5}
\end{equation}%
The combined action of all the transformations is given by%
\begin{equation}
U_{s}=\prod\limits_{\alpha =1}^{10}e^{iG_{\alpha }s_{\alpha }}  \label{3.6}
\end{equation}

The Galilei group admits a nontrivial central extension by a central charge
that commutes with all its generators. Such an extension is obtained as a
semi-direct product between the Galilei algebra $\mathcal{G}$ and the
algebra generated by the central charge, which in this case denotes the mass
operator $M=mI$, where $I$ is the identity operator and $m$ is the mass, $%
\mathcal{G}\times \langle M\rangle $ (see \cite{Weinberg}, \cite{Bose}). The
commutators corresponding to the extension are those of eqs.(\ref{3.1}),
with the exception of eq.($1_{f}$), which is replaced by

\begin{equation}
\lbrack G_{u_{i}},G_{rj}]=i\delta _{ij}M  \label{3.2}
\end{equation}%
While for an ordinary representation (or at the classical level) this
extension is unnecessary, for quantum representations with an arbitrary
phase (i.e., such that $\left\vert \phi \right\rangle \sim \exp \left( {%
i\omega }\right) \left\vert \phi \right\rangle $\textbf{\ }) it\ is
unavoidable (\cite{HW}, \cite{LB} Chapter 3). In this central extension, the
symmetry generators represent the basic magnitudes of the theory: the energy 
$H=\hbar G_{\tau }$, the three momentum components $P_{i}=\hbar G_{r_{i}}$,
the three angular momentum components $J_{i}=\hbar G_{\theta _{i}}$, and the
three Galilei-boost components $K_{i}^{(G)}=\hbar G_{u_{i}}$. The rest of
the physical magnitudes can be defined in terms of these basic ones: for
instance, the three position components are $Q_{i}=K_{i}^{(G)}/m$, the three
orbital angular momentum components are $L_{i}=\varepsilon _{ijk}Q^{j}P^{k}$%
, the three spin components are $S_{i}=J_{i}-L_{i}$. Then, by taking $\hbar
=1$, the commutation relations result%
\begin{equation}
\begin{tabular}{llll}
$\lbrack P_{i},P_{j}]=0$ & $(6_{a})$ \ \ \ \ \ \ \ \ \ \ \  & $%
[K_{i}^{(G)},P_{j}]=i\delta _{ij}M$ & $(6_{f})$ \\ 
$\lbrack K_{i}^{(G)},K_{j}^{(G)}]=0$ & $(6_{b})$ & $[P_{i},H]=0$ & $(6_{g})$
\\ 
$\lbrack J_{i},J_{j}]=i\varepsilon _{ijk}J^{k}$ & $(6_{c})$ & $[J_{i},H]=0$
& $(6_{h})$ \\ 
$\lbrack J_{i},P_{j}]=i\varepsilon _{ijk}P^{k}$ & $(6_{d})$ & $%
[K_{i}^{(G)},H]=iP_{i}$ & $(6_{i})$ \\ 
$\lbrack J_{i},K_{j}^{(G)}]=i\varepsilon _{ijk}K^{(G)k}$ & $(6_{e})$ &  & 
\end{tabular}
\label{3.3}
\end{equation}

Let us recall that a Casimir operator of a Lie group is an operator that
commutes with all the generators of the group and, therefore, is invariant
under all the transformations of the group. In the case of the mass
central-extended Galilei group, the Casimir operators are%
\begin{eqnarray}
C_{1}^{G} &=&M=mI  \label{3.71} \\
C_{2}^{G} &=&MH-P^{2}/2=M(H-P^{2}/2m)=mW  \label{3.72} \\
C_{4}^{G}
&=&M^{2}J_{i}J^{i}+(P_{i}P^{i})(K_{i}^{(G)}K^{(G)i})-(P_{i}K^{(G)i})^{2}-2MJ^{k}\varepsilon _{ijk}P^{i}K^{(G)j}
\label{3.73}
\end{eqnarray}%
where $W$ is the internal energy operator. In the reference frame at rest
with respect to the center of mass, these operators have the form 
\begin{equation}
C_{1}^{G}=M=mI\quad \;\qquad C_{2}^{G}=mW\quad \;\qquad
C_{4}^{G}=m^{2}J_{i}J^{i}=m^{2}S^{2}  \label{3.8}
\end{equation}%
The eigenvalues of the Casimir operators label the irreducible
representations of the group (\cite{wigner}, \cite{bargmann}, \cite{levy}).
So, in each irreducible representation, the Casimir operators are multiples
of the identity since $M=mI$, $W=wI$ (where $w$ is the internal energy), and 
$S^{2}=s(s+1)I$ (where $s$ is the eigenvalue of the spin) .

\section{Interpretation and Galilei group}

A continuous transformation, as in the case of the Galilei group, admits two
interpretations. Under the active interpretation, the transformation
corresponds to a change from one system to another $-$transformed$-$ system;
under the passive interpretation, the transformation consists in a change of
the viewpoint $-$reference frame$-$ from which the system is described (see 
\cite{Brading}). Nevertheless, in both cases the validity of a group of
symmetry transformations expresses the fact that the identity and the
behavior of the system are not altered by the application of the
transformations: in the active interpretation language, the original and the
transformed systems are equivalent; in the passive interpretation language,
the original and the transformed reference frames are equivalent. Then, any
realist interpretation should agree with that physical fact: the rule of
actual-value ascription should select a set of actual-valued observables
that remains unaltered under the transformations. Since the Casimir
operators of the central-extended Galilei group are invariant under all the
transformations of the group, one can reasonably expect that those Casimir
operators belong to the set of the actual-valued observables.

As we have seen, the preferred context selected by the Modal-Hamiltonian
actualization rule only depends on the Hamiltonian of the system. Then, the
requirement of invariance of the preferred context under the Galilei
transformations is directly fulfilled when the Hamiltonian is invariant,
that is, in the case of time-displacement, space-displacement and
space-rotation:%
\begin{eqnarray}
H^{\prime } &=&e^{iH\tau }H\,e^{-iH\tau }=H\ \text{\ (since }\left[ H,H%
\right] =0\text{)}  \nonumber \\
H^{\prime } &=&e^{iP_{i}r_{i}}H\,e^{-iP_{i}r_{i}}=H\ \ \text{(since }\left[
P_{i},H\right] =0\text{, see eq.}(6_{g})\text{)}  \nonumber \\
H^{\prime } &=&e^{iJ_{i}\theta _{i}}H\,e^{-iJ_{i}\theta _{i}}=H\ \ \text{%
(since }\left[ J_{i},H\right] =0\text{, see eq.}(6_{h})\text{)}  \label{4-1}
\end{eqnarray}%
However, it is not clear that the requirement of invariance of the preferred
context completely holds, since the Hamiltonian is not invariant under
Galilei-boosts. In fact, under a Galilei-boost transformation corresponding
to a velocity $u_{x}$, $H$ changes as%
\begin{equation}
H^{\prime }=e^{iK_{x}^{(G)}u_{x}}H\,e^{-iK_{x}^{(G)}u_{x}}\neq H\ \ \text{%
(since }\left[ K_{x}^{(G)},H\right] =iP_{x}\neq 0\text{, see eq.}(6_{i})%
\text{)}  \label{4-2}
\end{equation}%
Nevertheless, as we have shown in a previous work (\cite{ACL}), when space
is homogeneous and isotropic $-$when there are no external fields applied to
the system$-$, a Galilei-boost transformation only introduces a change in
the subsystem that carries the kinetic energy of translation: the internal
energy $W$ remains unaltered under the transformation. This should not sound
surprising to the extent that the internal energy $-$multiplied by $m-$ is a
Casimir operator of the central-extended Galilei group (see eq.(\ref{3.8})).

On this basis, we can reformulate the actualization rule in an explicit
Galilei-invariant form in terms of the Casimir operators of the
central-extended group:

\begin{quotation}
\textbf{Actualization rule'\ (AR'):} Given a quantum system free from
external fields and represented by $\mathcal{S}\mathrm{:}\;\mathrm{(}%
\mathcal{O}\,\mathrm{,}\,H\mathrm{)}$, its actual-valued observables are the
observables $C_{i}^{G}$ represented by the Casimir operators of the
central-extended Galilei group in the corresponding irreducible
representation, and all the observables commuting with the $C_{i}^{G}$ and
having, at least, the same symmetries as the $C_{i}^{G}$.
\end{quotation}

Since the Casimir operators of the central-extended Galilei group $-$in the
reference frame of the center of mass$-$ are $M$, $mW$ and $m^{2}S^{2}$,
this reformulation of the rule is in agreement with the original AR when
applied to a system free from external fields:

\begin{itemize}
\item The actual-valuedness of $M$ and $S^{2}$, postulated by AR', follows
from AR: these observables commute with $H$ and do not break its symmetries
because, in non-relativistic quantum mechanics, both are multiples of the
identity in any irreducible representation. The fact that $M$ and $S^{2}$
always acquire actual values is completely natural from a physical
viewpoint, since mass and spin are properties supposed to be always
possessed by any quantum system and measurable in any physical situation.

\item The actual-valuedness of $W$ might seem to be in conflict with AR
because $W$ is not the Hamiltonian: whereas $W$ is Galilei-invariant, $H$
changes under the action of a Galilei-boost. However, this is not a real
obstacle because a Galilei-boost transformation only introduces a change in
the subsystem that carries the kinetic energy of translation, which can be
considered a mere shift in an energy defined up to a constant (see \cite{ACL}%
).
\end{itemize}

Summing up, the application of the modal-Hamiltonian actualization rule
leads to reasonable results, since the actual-valued observables turn out to
be invariant and, therefore, objective magnitudes. The assumption of a
strong link between invariance and objectivity is rooted in a natural idea:
what is objective should not depend on the particular perspective used for
the description; or, in group-theoretical terms, what is objective according
to a theory is what is invariant under the symmetry group of the theory.
This idea is not new. It was widely discussed in the context of special and
general relativity with respect to the ontological status of space and time
(see \cite{Minkowski}). The claim that objectivity means invariance is also
a central thesis of Weyl's book \textit{Symmetry} (\cite{Weyl}). In recent
times, the idea has strongly reappeared in several works (\cite{Auyang}, 
\cite{Nozick}, \cite{Brading-Castellani} \cite{Earman-1}, \cite{Earman-2}).
From this perspective, the Modal-Hamiltonian actualization rule says that
the observables that acquire actual values are those representing objective
magnitudes. When expressed in so simple terms, we can expect that the rule
can be extrapolated to any quantum theory endowed with a symmetry group. In
particular, the actual-valued observables of a system in relativistic
quantum mechanics would be those represented by the Casimir operators of the
Poincar\'{e} group and of the internal symmetry group. In the following
sections we will develop this idea in detail.

\section{The Poincar\'{e} group and its limits}

In the case of the Poincar\'{e} group, the generators are $%
H,P_{i},J_{i,}K_{i}^{(P)}$, where the $K_{i}^{(P)}$ are the Lorentz-boost
components. The commutation relations of the Poincar\'{e} group can be
formulated in the 4-dimension Lorentz space-time as%
\begin{eqnarray}
\lbrack P_{\mu },P_{\nu }] &=&0\qquad \quad \lbrack M_{\mu \nu },P_{\rho
}]=\eta _{\mu \rho }P^{\nu }-\eta _{\nu \rho }P^{\mu }  \nonumber \\
\lbrack M_{\mu \nu },M_{\rho \sigma }] &=&\eta _{\mu \rho }M^{\nu \sigma
}-\eta _{\mu \sigma }M^{\nu \rho }-\eta _{\nu \rho }M^{\mu \rho }+\eta _{\nu
\sigma }M^{\mu \rho }  \label{5-1}
\end{eqnarray}%
where $\mu ,\nu ,...=0,1,2,3$, $\eta _{\mu \nu }$ is the metric tensor of
space-time, and%
\begin{equation}
P_{\mu }=(H,P_{i})\qquad M_{\mu \nu }=\left( 
\begin{array}{cc}
0 & K_{i}^{(P)} \\ 
-K_{i}^{(P)} & J_{ij}%
\end{array}%
\right) \qquad J_{k}=\varepsilon _{kij}J^{ij}  \label{5-2}
\end{equation}%
Then, eqs.(\ref{5-1}) can be rewritten in a form that permit them to be
compared with the Galilei case:%
\begin{equation}
\begin{tabular}{llll}
$\lbrack P_{i},P_{j}]=0$ & $(15_{a})$ \ \ \ \ \ \ \ \ \  & $%
[K_{i}^{(P)},P_{j}]=i\delta _{ij}H$ & $(15_{f})$ \\ 
$\lbrack K_{i}^{(P)},K_{j}^{(P)}]=-i\varepsilon _{ijk}J^{k}$ & $(15_{b})$ & $%
[P_{i},H]=0$ & $(15_{g})$ \\ 
$\lbrack J_{i},J_{j}]=i\varepsilon _{ijk}J^{k}$ & $(15_{c})$ & $[J_{i},H]=0$
& $(15_{h})$ \\ 
$\lbrack J_{i},P_{j}]=i\varepsilon _{ijk}P^{k}$ & $(15_{d})$ & $%
[K_{i}^{(P)},H]=iP_{i}$ & $(15_{i})$ \\ 
$\lbrack J_{i},K_{j}^{(P)}]=i\varepsilon _{ijk}K^{(P)k}$ & $(15_{e})$ &  & 
\end{tabular}
\label{5-3}
\end{equation}%
In turn, the Casimir operators of the Poincar\'{e} group are (see \cite{LL}) 
\begin{equation}
\begin{array}{l}
C_{2}^{P}=H^{2}-P_{i}P^{i} \\ 
C_{4}^{P}=H^{2}J_{i}J^{i}+(P_{i}P^{i})(K_{i}^{(P)}K^{(P)i})-(J_{i}P^{i})^{2}-(P_{i}K^{(P)i})^{2}-2HJ^{k}\varepsilon _{ijk}P^{i}K^{(P)j}%
\end{array}
\label{5-4}
\end{equation}%
In the reference frame at rest with respect to the center of mass, where $%
P_{i}=0$ and $H=E=m_{0}$, these operators result 
\begin{equation}
C_{2}^{P}=m_{0}^{2}I\qquad \;\qquad
C_{4}^{P}=m_{0}^{2}J_{i}J^{i}=m_{0}^{2}S^{2}  \label{5-5}
\end{equation}

If we now extrapolate the invariant Modal-Hamiltonian actualization rule AR'
to the relativistic case, we have to conclude that the Casimir operators $%
C_{2}^{P}$ and $C_{4}^{P}$ are the operators that define the actual-valued
observables of the quantum system. This result is physically reasonable
because mass and spin are properties supposed to be always possessed by any
elemental particle (see \cite{Haag}); moreover, mass and spin are two of the
properties that contribute to the classification of elemental particles.
However, the adequacy of the interpretation in the relativistic realm is not
guaranteed yet, since it is still necessary to prove that the actual-valued
observables in the relativistic and in the non-relativistic theories are
correctly related through an adequate limit. This task leads us to analyze
the relationship between the Galilei group and the Poincar\'{e} group.

As it is well known, the Galilei group can be recovered from the Poincar\'{e}
group by means of an In\"{o}n\"{u}-Wigner contraction (see \cite{LL}).
However, as we have seen, the physically meaningful group of quantum
mechanics is not the Galilei group, but its central extension. Therefore,
the question is whether the central-extended Galilei group can be obtained
from a central extension of the Poincar\'{e} group. But the answer to this
question is not straightforward, because the Poincar\'{e} group does not
admit non-trivial central extensions (\cite{Car}).\footnote{%
A trivial extension of a Lie algebra $\mathfrak{g}$ is a direct sum $%
\mathfrak{g}\oplus M$, where $M$ is an additional commuting generator.} For
this reason, in the following subsections we will consider two limiting
procedures. First, we will review the traditional non-relativistic limit,
which has a clear physical meaning but does not admit a direct
representation in group terms. Then, we will introduce a generalized In\"{o}n%
\"{u}-Wigner contraction of a trivial extension of the Poincar\'{e} group,
which, as it will be shown, leads to the central-extended Galilei group.

\subsection{The traditional non-relativistic limit}

Let us recall the relativistic transformations of coordinates:%
\begin{eqnarray}
\overrightarrow{x}^{\prime } &=&R\overrightarrow{x}+\frac{(\gamma -1)}{%
\overrightarrow{\beta }^{2}}(\overrightarrow{\beta }\cdot \overrightarrow{x})%
\overrightarrow{\beta }-\gamma ct\overrightarrow{\beta }+\overrightarrow{r}
\label{5-6} \\
ct^{\prime } &=&\gamma (ct-\overrightarrow{\beta }\cdot \overrightarrow{x}%
)+c\tau  \label{5-7}
\end{eqnarray}%
where $\overrightarrow{r}$ is the space-displacement vector, $\tau $ is the
time-displacement scalar, $R$ is the space-rotation matrix, and $\gamma =(1-%
\overrightarrow{\beta }^{2})^{-1/2}$ with $\overrightarrow{\beta }=%
\overrightarrow{v}/c$. These are the transformations that lead to the Poincar%
\'{e} group given by eqs.(\ref{5-3}), where the parameters corresponding to
each generator are: $\tau $ for $H$, $\overrightarrow{r}$ for the $P_{i}$, $%
R $ for the $J_{i}$, and $\overrightarrow{\beta }$ for the $K_{i}^{(P)}$.

The traditional relativistic limit is the limit $\beta \rightarrow 0$ ($%
\gamma \rightarrow 1$). This means that the limit affects only the
boost-transformation, and not the remaining transformations. This fact can
also be noted by comparing the central-extended Galilei group in eqs.(\ref%
{3.3}) with the Poincar\'{e} group in eqs.(\ref{5-3}): the two groups share
a splittable seven dimensional subgroup $ISO(3)\times \left\langle
H\right\rangle $, defined by the commutation relations $(15_{a})$, $(15_{c})$%
, $(15_{d})$, $(15_{g})$ and $(15_{h})$. In particular, $\left\langle
H\right\rangle $ is the time-displacement group generated by $H$, and $%
ISO(3)=\left\langle T_{i}\right\rangle \times SO(3)$ is the inhomogeneous
rotation group in three dimensions, where $\left\langle T_{i}\right\rangle $
is the space-displacement group generated by the $P_{i}$ and $SO(3)$ is the
space-rotation group generated by the $J_{i}$. Therefore, the difference
between the Galilei and the Poincar\'{e} groups is confined to the
commutation relations that involve the boost generators: the relativistic
limit should turn the Poincar\'{e} boost-generators $K_{i}^{(P)}$ into the
Galilei boost-generators $K_{i}^{(G)}$, and the commutation relations $%
(15_{b})$, $(15_{e})$, $(15_{f})$ and $(15_{i})$ of the Poincar\'{e} group
into the commutation relations $(6_{b})$, $(6_{e})$, $(6_{f})$ and $(6_{i})$
of the Galilei group respectively.

Since in this case we are interested only in boosts, we can simplify the
transformations of coordinates of eqs.(\ref{5-6}) and (\ref{5-7}) by making $%
\tau =0$, $\overrightarrow{r}=\overrightarrow{0}$ and $R=I$:%
\begin{eqnarray}
\overrightarrow{x}^{\prime } &=&\overrightarrow{x}+\frac{(\gamma -1)}{%
\overrightarrow{\beta }^{2}}(\overrightarrow{\beta }\cdot \overrightarrow{x})%
\overrightarrow{\beta }-\gamma ct\overrightarrow{\beta }  \label{5-8} \\
ct^{\prime } &=&\gamma (ct-\overrightarrow{\beta }\cdot \overrightarrow{x})
\label{5-9}
\end{eqnarray}%
Let us also consider that energy, mass and momentum are%
\begin{equation}
E=\gamma m_{0}c^{2}\qquad \qquad m=\gamma m_{0}\qquad \qquad p_{i}=\gamma
m_{0}v_{i}  \label{5-10}
\end{equation}%
As it is well known, in the traditional relativistic limit $\beta
\rightarrow 0$ ($\gamma \rightarrow 1$) we obtain%
\begin{eqnarray}
\overrightarrow{x}^{\prime } &=&\overrightarrow{x}\qquad \qquad \qquad
t^{\prime }=t  \label{5-11} \\
E &=&m_{0}c^{2}\qquad \qquad m=m_{0}\qquad \qquad p_{i}=m_{0}v_{i}
\label{5-12}
\end{eqnarray}%
On the other hand, the Poincar\'{e}-boost generators $K_{i}^{(P)}$ can be
expressed as%
\begin{equation}
K_{i}^{(P)}=X_{i}H-X_{0}P_{i}  \label{5-13}
\end{equation}%
where the $X_{i}$ are the position operators corresponding to the $x_{i}$, $%
H $ is the Hamiltonian operator corresponding to the energy $E$, $X_{0}$ is
the operator conjugate to $H$ and, then, it corresponds to $ct$, and the $%
P_{i}$ are the momentum operators corresponding to the $p_{i}$. Therefore,
by considering eqs.(\ref{5-11}) and (\ref{5-12}), the relativistic limit of
the $K_{i}^{(P)}$ is%
\begin{equation}
\lim_{\beta \rightarrow 0}K_{i}^{(P)}=X_{i}m_{0}c^{2}-ctP_{i}=K_{i}^{(\beta
\rightarrow 0)}  \label{5-14}
\end{equation}%
Now, we can compute the commutation relations $(15_{b})$, $(15_{e})$, $%
(15_{f})$ and $(15_{i})$ with the just obtained $K_{i}^{(\beta \rightarrow
0)}$:%
\begin{eqnarray}
\lim_{\beta \rightarrow 0}\left[ K_{i}^{(P)},K_{j}^{(P)}\right] &=&\left[
K_{i}^{(\beta \rightarrow 0)},K_{j}^{(\beta \rightarrow 0)}\right] =0
\label{5-15} \\
\lim_{\beta \rightarrow 0}\left[ J_{i}^{(P)},K_{j}^{(P)}\right] &=&\left[
J_{i}^{(P)},K_{j}^{(\beta \rightarrow 0)}\right] =i\varepsilon
_{ijk}K^{(\beta \rightarrow 0)k}  \label{5-16} \\
\lim_{\beta \rightarrow 0}\left[ K_{i}^{(P)},P_{j}\right] &=&\left[
K_{i}^{(\beta \rightarrow 0)},P_{j}\right] =i\delta _{ij}M_{0}c^{2}
\label{5-17} \\
\lim_{\beta \rightarrow 0}\left[ K_{i}^{(P)},H\right] &=&\left[
K_{i}^{(\beta \rightarrow 0)},H\right] =iP_{i}  \label{5-18}
\end{eqnarray}%
In turn, by making $c=1$, eq.(\ref{5-17}) becomes 
\begin{equation}
\lim_{\beta \rightarrow 0}\left[ K_{i}^{(P)},P_{j}\right] =\left[
K_{i}^{(\beta \rightarrow 0)},P_{j}\right] =i\delta _{ij}M_{0}  \label{5-19}
\end{equation}%
Finally, let us compare eqs.(\ref{5-15}), (\ref{5-16}), (\ref{5-19}), and (%
\ref{5-18}) with the corresponding commutation relations $(6_{b})$, $(6_{e})$%
, $(6_{f})$ and $(6_{i})$ of the Galilei group: if the limit $K_{i}^{(\beta
\rightarrow 0)}$ of the Poincar\'{e}-boost is identified with the
Galilei-boost $K_{i}^{(G)}$, and the Poincar\'{e} operator $M_{0}$ is
identified with the Galilei mass operator $M$, the central-extended Galilei
group can be considered the non-relativistic limit of the Poincar\'{e} group.

\subsection{A generalized In\"{o}n\"{u}-Wigner contraction}

The traditional non-relativistic limit has a clear physical meaning and,
then, it is desirable to express it in group terms. We know that the
traditional In\"{o}n\"{u}-Wigner contraction maps the Poincar\'{e} group
onto the Galilei group. But, since the mass generator has been added to the
Galilei group, an analogous map between the Poincar\'{e} and the
central-extended Galilei groups is not possible, to the extent that both
groups have different numbers of generators. Therefore, a natural way of
obtaining the desired map is by extending the Poincar\'{e} group. This is
the strategy that we will follow below.

Let us consider the Poincar\'{e} group $ISO(1,3)$, with generators $%
\{H,P_{i},J_{i},K_{i}^{(P)}\}$, and its corresponding commutation relations
given by eqs.(\ref{5-3}). Let us recall that the Poincar\'{e} group does not
admit non-trivial extensions. Therefore, we extend the group trivially, in
such a way that all the generators commute with a central charge $M$. In
this case, we obtain a new group $ISO(1,3)\times \left\langle M\right\rangle 
$, with generators $\{H,P_{i},J_{i},K_{i}^{(P)},M\}$, corresponding to a 
\textit{trivially extended Poincar\'{e} group}. Now, we can introduce the
following change in the basis of generators:%
\begin{equation}
\overline{H}=H-M  \label{5-20}
\end{equation}%
In the new basis $\{\overline{H},P_{i},J_{i},K_{i}^{(P)},M\}$, the
commutation relations given by eqs.(\ref{5-3}) preserve their form, with the
only exception of eq.$(15_{f})$, which becomes:%
\begin{equation}
\left[ K_{i}^{(P)},P_{j}\right] =i\delta _{ij}H=i\delta _{ij}(\overline{H}+M)
\label{5-21}
\end{equation}

Now the task is to show that this trivially extended Poincar\'{e} group
represented by $ISO(1,3)\times \left\langle M\right\rangle $ contracts to
the centrally extended Galilei group $\mathcal{G}\times \left\langle
M\right\rangle $:%
\begin{equation}
ISO(1,3)\times \left\langle M\right\rangle \longrightarrow \mathcal{G}\times
\left\langle M\right\rangle  \label{5-22}
\end{equation}%
The contraction is obtained by rescaling the generators as%
\begin{equation}
J_{i}^{\prime }=J_{i}\qquad P_{i}^{\prime }=\varepsilon P_{i}\qquad
K_{i}^{(P)\prime }=\varepsilon K_{i}^{(P)}\qquad \overline{H}^{\prime }=%
\overline{H}\qquad M^{\prime }=\varepsilon ^{2}M  \label{5-23}
\end{equation}%
The commutation relations given by eqs.(\ref{5-3}) are left unchanged by the
rescaling, with the exception of eq.$(15_{b})$, and of eq.$(15_{f})$ now
replaced by eq.(\ref{5-21}):%
\begin{eqnarray}
\left[ K_{i}^{(P)\prime },K_{j}^{(P)\prime }\right] &=&i\varepsilon
_{ijk}\varepsilon ^{2}J^{\prime k}  \label{5-24} \\
\left[ K_{i}^{(P)\prime },P_{j}^{\prime }\right] &=&i\delta
_{ij}(\varepsilon ^{2}\overline{H}^{\prime }+M^{\prime })  \label{5-25}
\end{eqnarray}%
As in the original In\"{o}n\"{u}-Wigner contraction, the operation is
completed by introducing the limit $\varepsilon \rightarrow 0$, which turns
eqs.(\ref{5-24}) and (\ref{5-25}) into%
\begin{equation}
\left[ K_{i}^{(P)\prime },K_{j}^{(P)\prime }\right] =0\qquad \qquad \left[
K_{i}^{(P)\prime },P_{j}^{\prime }\right] =0  \label{5-26}
\end{equation}

The In\"{o}n\"{u}-Wigner contraction admits a physical interpretation (see 
\cite{LL}). The factor $\varepsilon $ affects the boost generators $%
K_{i}^{(P)\prime }$and the momentum generators $P_{i}^{\prime }$. As a
consequence, $\varepsilon $ also affects the boost-velocities and the
space-displacements resulting from the exponentiation of those generators.
Then, by introducing the limit $\varepsilon \rightarrow 0$, we describe a
situation where boost-velocities and space-displacements are
\textquotedblleft small\textquotedblright . Boost-velocities are small with
respect to the velocity of light $c$, which here was taken as $c=1$.
Space-displacements are small with respect to $c\tau $, where $\tau $ is the
time-displacement associated with the Hamiltonian $H$, which is not affected
by $\varepsilon $. \ For these reasons, this kind of contraction is known as
\textquotedblleft speed-space contraction\textquotedblright\ (\cite{LL}).

Summing up, the result of the application of this generalized In\"{o}n\"{u}%
-Wigner contraction to the trivially extended Poincar\'{e} group is

\begin{equation}
\begin{tabular}{llll}
$\lbrack P_{i}^{\prime },P_{j}]=0$ & $(39_{a})$ \ \ \ \ \ \ \ \ \  & $%
[K_{i}^{(P)\prime },P_{j}^{\prime }]=i\delta _{ij}M^{\prime }$ & $(39_{f})$
\\ 
$\lbrack K_{i}^{(P)\prime },K_{j}^{(P)\prime }]=0$ & $(39_{b})$ & $%
[P_{i}^{\prime },\overline{H}^{\prime }]=0$ & $(39_{g})$ \\ 
$\lbrack J_{i}^{\prime },J_{j}^{\prime }]=i\varepsilon _{ijk}J^{k\prime }$ & 
$(39_{c})$ & $[J_{i}^{\prime },\overline{H}^{\prime }]=0$ & $(39_{h})$ \\ 
$\lbrack J_{i}^{\prime },P_{j}^{\prime }]=i\varepsilon _{ijk}P^{k\prime }$ & 
$(39_{d})$ & $[K_{i}^{(P)\prime },\overline{H}^{\prime }]=iP_{i}^{\prime }$
& $(39_{i})$ \\ 
$\lbrack J_{i}^{\prime },K_{j}^{(P)\prime }]=i\varepsilon
_{ijk}K^{(P)k\prime }$ & $(39_{e})$ &  & 
\end{tabular}
\label{5-27}
\end{equation}%
Let us compare these eqs.(\ref{5-27}) with the commutation relations given
by eqs.(\ref{3.3}), which define the central-extended Galilei group. If the
mass $M^{\prime }$ of relation $(39_{f})$ is identified with the mass $M$ of
relation $(6_{f})$, and the Poincar\'{e}-boost $K_{i}^{(P)\prime }$ of eqs.(%
\ref{5-27}) is identified with the Galilei-boost $K_{i}^{(G)}$ of eqs.(\ref%
{3.3}), then it can be said that the generalized In\"{o}n\"{u}-Wigner
contraction of the trivially extended Poincar\'{e} group leads to the
central-extended Galilei group, as originally expected (see \cite{Rutwig}).

\section{The limits of the Casimir operators}

Let us recall that the physically meaningful group of non-relativistic
quantum mechanics is not the Galilei group, but its central extension, whose
Casimir operators, expressed in the reference frame of the center of mass,
are (see eqs.(\ref{3.8}))%
\begin{eqnarray}
C_{1}^{G} &=&M=mI\quad  \label{6-1} \\
\;\qquad C_{2}^{G} &=&mW=mwI  \label{6-2} \\
C_{4}^{G} &=&m^{2}J_{i}J^{i}=m^{2}S^{2}=m^{2}s(s+1)I  \label{6-3}
\end{eqnarray}%
In turn, the Casimir operators of the Poincar\'{e} group, expressed in the
reference frame of the center of mass, are (see eqs.(\ref{5-5}))%
\begin{eqnarray}
C_{2}^{P} &=&m_{0}^{2}I  \label{6-4} \\
\qquad \;C_{4}^{P} &=&m_{0}^{2}J_{i}J^{i}=m_{0}^{2}S^{2}=m_{0}^{2}s(s+1)I
\label{6-5}
\end{eqnarray}%
It is quite clear that there is no limit that can introduce a map between
the two $C_{i}^{P}$ and the three $C_{j}^{G}$. Nevertheless, in the
traditional relativistic limit $\beta \rightarrow 0$ ($\gamma \rightarrow 1$%
), $m=\gamma m_{0}$ becomes $m_{0}$ and $E=\gamma m_{0}c^{2}$ becomes $%
m_{0}c^{2}$. Therefore, by making $c=1$, in the non-relativistic limit, both
the mass $m$ and the internal energy $E=w$ are $m_{0}$. This means that,
conceptually, the limit of $C_{4}^{P}$ is $C_{4}^{G}$, but the limit of $%
C_{2}^{P}$ leads to the two remaining Casimir operators $C_{1}^{G}$ and $%
C_{2}^{G}$, since in this limit $m=w=m_{0}$ and, thus, $%
C_{2}^{G}=(C_{1}^{G})^{2}$:%
\begin{equation}
C_{4}^{P}\longrightarrow C_{4}^{G}\qquad \qquad C_{2}^{P}\longrightarrow
C_{2}^{G}=\left( C_{1}^{G}\right) ^{2}  \label{6-6}
\end{equation}%
Of course, this is a conceptual argument that cannot be expressed in group
language, to the extent that the limit relates a non-extended group with an
extended group. Then, we may expect that, by following the strategy
developed in the previous section, the generalized In\"{o}n\"{u}-Wigner
contraction of the Casimir operators of the trivially extended Poincar\'{e}
group leads to the Casimir operators of the central-extended Galilei group.

The Casimir operators of the trivially extended Poincar\'{e} group
represented by $ISO(1,3)\times \left\langle M\right\rangle $ in the basis $\{%
\overline{H},P_{i},J_{i},K_{i}^{(P)},M\}$ are%
\begin{eqnarray}
C_{1}^{PE} &=&M=mI\quad   \label{6-7} \\
\;\qquad C_{2}^{PE} &=&-(P_{i}P^{i})+\overline{H}^{2}+M^{2}+2\overline{H}M
\label{6-8} \\
C_{4}^{PE} &=&\left( \overline{H}+M\right) ^{2}J_{i}J^{i}-\left(
J_{i}P^{i}\right) ^{2}+\left( P_{i}P^{i}\right) \left(
K_{i}^{(P)}K^{(P)i}\right) -  \label{6-9} \\
&&-\left( P_{i}K^{(P)i}\right) ^{2}-2\left( \overline{H}+M\right)
J^{k}\varepsilon _{ijk}P^{i}K^{(P)j}  \nonumber
\end{eqnarray}%
By means of the rescaled basis introduced in eqs.(\ref{5-23}), the Casimir
operators are%
\begin{eqnarray}
\widetilde{C}_{1}^{PE} &=&\varepsilon ^{-2}M^{\prime }\quad   \label{6-10} \\
\;\qquad \widetilde{C}_{2}^{PE} &=&-\varepsilon ^{-2}(P_{i}^{\prime
}P^{\prime i})+\overline{H}^{\prime 2}+\varepsilon ^{-4}M^{\prime
2}+2\varepsilon ^{-2}\overline{H}^{\prime }M^{\prime }  \label{6-11} \\
C_{4}^{PE} &=&(\overline{H}^{\prime }+\varepsilon ^{-2}M^{\prime
})^{2}J_{i}^{\prime }J^{\prime i}-\varepsilon ^{-2}(J_{i}^{\prime }P^{\prime
i})^{2}+\varepsilon ^{-4}(P_{i}^{\prime }P^{\prime i})(K_{i}^{(P)\prime
}K^{(P)\prime i})-  \label{6-12} \\
&&-\varepsilon ^{-4}(P_{i}^{\prime }K^{(P)\prime i})^{2}-2\varepsilon
^{-2}\left( \overline{H}^{\prime }+\varepsilon ^{-2}M^{\prime }\right)
J^{\prime k}\varepsilon _{ijk}P^{\prime i}K^{(P)\prime j}  \nonumber
\end{eqnarray}%
As usual, the contracted Casimir operators are obtained by applying the
limit $\varepsilon \rightarrow 0$ to the adequately rescaled operators:%
\begin{eqnarray}
\widehat{C}_{1}^{PE} &=&\lim_{\varepsilon \rightarrow 0}\varepsilon ^{2}%
\widetilde{C}_{1}^{PE}=M^{\prime }\quad   \label{6-13} \\
\;\qquad \widehat{C}_{2}^{PE} &=&\lim_{\varepsilon \rightarrow 0}\varepsilon
^{4}\widetilde{C}_{2}^{PE}=M^{\prime 2}  \label{6-14} \\
\widehat{C}_{4}^{PE} &=&\lim_{\varepsilon \rightarrow 0}\varepsilon
^{4}C_{4}^{PE}=M^{\prime }{}^{2}J_{i}^{\prime }J^{\prime i}+(P_{i}^{\prime
}P^{\prime i})(K_{i}^{(P)\prime }K^{(P)\prime i})-  \label{6-15} \\
&&-(P_{i}^{\prime }K^{(P)\prime i})^{2}-2M^{\prime }J^{\prime k}\varepsilon
_{ijk}P^{\prime i}K^{(P)\prime j}  \nonumber
\end{eqnarray}%
Let us compare these eqs.(\ref{6-13}), (\ref{6-14}) and (\ref{6-15}) with
eqs.(\ref{3.71}), (\ref{3.72}) and (\ref{3.73}), which express the Casimir
operators of the mass central-extended Galilei group in the reference frame
at rest with respect to the center of mass. As in the case of the
commutation relations, if the mass $M^{\prime }$ of the first group of
equations is identified with the mass $M$ of the second group, and the
Poincar\'{e}-boost $K^{(P)\prime i}$ is identified with the Galilei-boost $%
K_{i}^{(G)}$, it can be said that the generalized In\"{o}n\"{u}-Wigner
contraction of the Casimir operators of the trivially extended Poincar\'{e}
group leads to the Casimir operators of the central-extended Galilei group.

Summing up, when the Modal-Hamiltonian actualization rule is expressed in an
explicit Galilei-invariant form, it leads to a physically reasonable result:
the actual-valued observables are those represented by the Casimir operators
of the mass central-extended Galilei group, $M$, $W$ and $S^{2}$, which
acquire their actual values $m$, $w$ and $s(s+1)$. The natural strategy is
to extrapolate the interpretation to the relativistic realm by replacing the
Galilei group with the Poincar\'{e} group. But when one takes into account
that the relevant group of non-relativistic quantum mechanics is not the
Galilei group but its central extension, the mere replacement of the
relevant group is not sufficient: one has to show also that the
actual-valued observables in the relativistic and the non-relativistic cases
are related through the adequate limit. As a consequence, the Poincar\'{e}
group has to be trivially extended, in order to show that the limit between
the corresponding Casimir operators holds, and this result counts in favor
of the proposed extrapolation of our MHI to relativistic quantum mechanics.

\section{Relativistic quantum mechanics}

Since the spirit of the MHI is to consider the observables representing
invariances as the actual-valued observables of the system, when this
interpretation is extrapolated to the relativistic domain, all the
symmetries have to be considered. In particular, in relativistic quantum
theories, besides the space-time symmetries represented by the Poincar\'{e}
group, quantum systems have internal gauge-symmetries. Therefore, according
to the MHI, the invariant magnitudes corresponding to those gauge-symmetries
should also be actual-valued. As an illustration of this claim, in this
section we will analyze the case of relativistic quantum mechanics with
gauge $U(1)$ fields.

\subsection{Internal symmetry}

Let us consider a free Dirac field $\Psi $ whose Lagrangian has the
following form: 
\begin{equation}
L_{D}=\overline{\Psi }\left( {i\hbar \partial _{t}-c\overrightarrow{\alpha }%
\cdot \overrightarrow{p}-\beta m_{o}c^{2}}\right) \Psi  \label{7.1}
\end{equation}%
where $\overrightarrow{\alpha }$ and $\beta $ are the Dirac matrices, and $%
\Psi $ is a four component spinor that is composed of two spinors. This
means that the field is $\Psi =\binom{\phi }{\chi }$ (and the conjugate
transposed is $\Psi ^{\dag }=\left( 
\begin{array}{cc}
\phi ^{\dag } & \chi ^{\dag }%
\end{array}%
\right) $ ), where $\phi =\binom{\phi _{1}}{\phi _{2}}$ and $\chi =\binom{%
\chi _{1}}{\chi _{2}}$. So, we can write the Lagrangian of eq.(\ref{7.1})
explicitly in terms of this spinors as%
\begin{equation}
L_{D}(\phi ,\chi )=\left( 
\begin{array}{cc}
\phi ^{\dag } & \chi ^{\dag }%
\end{array}%
\right) \left[ i\hbar \binom{\partial _{t}\phi }{\partial _{t}\chi }-c\left( 
\begin{array}{cc}
0 & \overrightarrow{\sigma } \\ 
\overrightarrow{\sigma } & 0%
\end{array}%
\right) \cdot \overrightarrow{p}\binom{\phi }{\chi }-m_{o}c^{2}\left( 
\begin{array}{cc}
I & 0 \\ 
0 & -I%
\end{array}%
\right) \binom{\phi }{\chi }\right]  \label{7.2}
\end{equation}%
where $\overrightarrow{\sigma }$ are the Pauli matrices. By computing the
inner product, eq.(\ref{7.2}) reads%
\begin{equation}
L_{D}(\phi ,\chi )=i\hbar \phi ^{\dag }\partial _{t}\phi +i\hbar \chi ^{\dag
}\partial _{t}\chi -\phi ^{\dag }c\overrightarrow{\sigma }\cdot 
\overrightarrow{p}\chi -\chi ^{\dag }c\overrightarrow{\sigma }\cdot 
\overrightarrow{p}\phi -m_{o}c^{2}\phi ^{\dag }\phi +m_{o}c^{2}\chi ^{\dag
}\chi  \label{7.3}
\end{equation}%
This Lagrangian is invariant under a global gauge-symmetry represented by
the Abelian Lie group $U(1)$, such that the field transforms as%
\begin{equation}
\Psi \rightarrow e^{-iQ\alpha }\Psi \qquad \qquad \overline{\Psi }%
\rightarrow \overline{\Psi }e^{iQ\alpha }  \label{7-2}
\end{equation}%
where $Q$ is the generator of the transformation and $\alpha $ is a constant
real number. As it is well known, $L_{D}$ is not invariant under a local
gauge-symmetry $U(1)$ that transforms the field as 
\begin{equation}
\Psi \rightarrow e^{-iQ\alpha \left( x\right) }\Psi \qquad \qquad \overline{%
\Psi }\rightarrow \overline{\Psi }e^{iQ\alpha \left( x\right) }  \label{7-3}
\end{equation}%
where $\alpha (x)$ is now a real function of the space-time position $x$. In
order to recover invariance, a field ${A_{\mu }}$ has to be included, such
that it is transformed as 
\begin{equation}
A_{\mu }\rightarrow A_{\mu }+\partial _{\mu }\alpha  \label{7-5}
\end{equation}%
and%
\begin{equation}
L_{M}=\frac{1}{4}F^{\mu \nu }F_{\mu \nu }\qquad \qquad F_{\mu \nu }=\partial
_{\mu }A_{\nu }-\partial _{\nu }A_{\mu }  \label{7-6}
\end{equation}%
Then, the invariant Lagrangian is%
\begin{equation}
L_{D_{f}}=\overline{\Psi }\left( {i\hbar (\partial _{t}-ieA_{o})-c%
\overrightarrow{\alpha }\cdot (\overrightarrow{p}-e\overrightarrow{A})-\beta
m_{o}c^{2}}\right) \Psi +L_{M}  \label{7-7}
\end{equation}%
In this case, $Q$ is trivially the only Casimir operator $C_{1}^{U}$ of the
internal group $U(1)$.

Since the internal gauge-symmetry $U(1)$ is a symmetry of the theory,
according to the MHI extrapolated to this case, the only Casimir operator $%
C_{1}^{U}=Q$ of this symmetry group $-$invariant under the corresponding
transformations$-$ is an actual valued observable of the system. Again, this
leads to a physically reasonable result since the operator $Q$ of the
internal gauge-symmetry $U(1)$ is the charge operator, $Q=eI$.

\subsection{The limit of the internal gauge-symmetry}

In the literature, it is usual to find the non-relativistic limit of the
Euler-Lagrange equations, but not of the Lagrangian. In order the obtain
this limit, we can introduce the following ansatz:

\begin{equation}
\binom{\phi }{\chi }=e^{-\frac{im_{o}c^{2}t}{\hbar }}\binom{\phi _{o}}{\chi
_{o}}  \label{7.4}
\end{equation}%
where $\phi _{o}$ and $\chi _{o}$ still depend on space and time
coordinates. Eq.(\ref{7.4}) expresses the spinors in terms of two separate
time-dependent factors: one unknown, given by $\phi _{o}$ and $\chi _{o}$,
and the other an oscillating factor with frequency $\omega _{o}=\frac{%
m_{o}c^{2}}{\hbar }$. By introducing eq.(\ref{7.4}) into eq.(\ref{7.3}), we
obtain%
\begin{equation}
L_{D}(\phi _{o},\chi _{o})=i\hbar \phi _{o}^{\dag }\partial _{t}\phi
_{o}+i\hbar \chi _{o}^{\dag }\partial _{t}\chi _{o}-\phi _{o}^{\dag }c%
\overrightarrow{\sigma }\cdot \overrightarrow{p}\chi _{o}-\chi _{o}^{\dag }c%
\overrightarrow{\sigma }\cdot \overrightarrow{p}\phi _{o}+2m_{o}c^{2}\chi
_{o}^{\dag }\chi _{o}  \label{7.5}
\end{equation}%
which can be rearranged as%
\begin{equation}
L_{D}(\phi _{o},\chi _{o})=i\hbar \phi _{o}^{\dag }\partial _{t}\phi
_{o}-\phi _{o}^{\dag }c\overrightarrow{\sigma }\cdot \overrightarrow{p}\chi
_{o}-\chi _{o}^{\dag }c\overrightarrow{\sigma }\cdot \overrightarrow{p}\phi
_{o}+i\hbar \chi _{o}^{\dag }(\partial _{t}+2m_{o}c^{2})\chi _{o}
\label{7.7}
\end{equation}%
The time-derivatives of the spinors $\phi _{o}$ and $\chi _{o}$ are related
with the time-oscillation with frequency $\omega =E_{o}/\hbar $.

Up to this point, no non-relativistic limit has been introduced yet. In
order to perform such a limit, we have to notice that the total energy for
the spinor $\phi _{o}$ is $E_{\phi _{o}}=E_{o}$ and for the spinor $\chi
_{o} $ is $E_{\chi _{o}}=E_{o}+2m_{o}c^{2}$. So, if we consider that $%
E_{o}\ll m_{o}c^{2}$, then%
\begin{equation}
E_{\phi _{o}}=E_{o}\ \ \ \ \ \ \ \ E_{\chi _{o}}=E_{o}+2m_{o}c^{2}\sim
2m_{o}c^{2}  \label{7&6}
\end{equation}%
These two relations imply that the last two terms of eq.(\ref{7.7}) can be
written as%
\begin{equation}
i\hbar \chi _{o}^{\dag }(\partial _{t}+2m_{o}c^{2})\chi _{o}=i\hbar \chi
_{o}^{\dag }(E+2m_{o}c^{2})\chi _{o}\sim i\hbar \chi _{o}^{\dag
}2m_{o}c^{2}\chi _{o}  \label{7.8}
\end{equation}%
and the non-relativistic limit of the Lagrangian reads%
\begin{equation}
\widetilde{L}_{D}(\phi _{o},\chi _{o})=i\hbar \phi _{o}^{\dag }\partial
_{t}\phi _{o}-\phi _{o}^{\dag }c\overrightarrow{\sigma }\cdot 
\overrightarrow{p}\chi _{o}-\chi _{o}^{\dag }c\overrightarrow{\sigma }\cdot 
\overrightarrow{p}\phi _{o}+i\hbar 2m_{o}c^{2}\chi _{o}^{\dag }\chi _{o}
\label{7.9}
\end{equation}

In order to write the Lagrangian of eq.(\ref{7.9}) in terms of only one of
the spinors, say $\phi _{o}$, we have to begin by computing the
Euler-Lagrange equation for $\chi _{o}$,

\begin{equation}
\partial _{t}(\frac{\partial \widetilde{L}_{D}}{\partial (\partial _{t}\chi
_{o}^{\dag })})+\partial _{i}(\frac{\partial \widetilde{L}_{D}}{\partial
(\partial _{i}\chi _{o}^{\dag })})-\frac{\partial \widetilde{L}_{D}}{%
\partial \chi _{o}^{\dag }}=0  \label{7.10}
\end{equation}%
which results

\begin{equation}
c\overrightarrow{\sigma }\cdot \overrightarrow{p}\phi _{o}-2m_{o}c^{2}\chi
_{o}=0  \label{7.11}
\end{equation}%
or, equivalently,

\begin{equation}
\chi _{o}=\frac{\overrightarrow{\sigma }\cdot \overrightarrow{p}\phi _{o}}{%
2m_{o}c}  \label{7.12}
\end{equation}%
If we now replace eq.(\ref{7.12}) into eq.(\ref{7.9}), we obtain%
\begin{eqnarray}
\widetilde{L}_{D}(\phi _{o},\chi _{o}) &=&i\hbar \phi _{o}^{\dag }\partial
_{t}\phi _{o}-\phi _{o}^{\dag }c\overrightarrow{\sigma }\cdot 
\overrightarrow{p}(\frac{\overrightarrow{\sigma }\cdot \overrightarrow{p}%
\phi _{o}}{2m_{o}c})-  \label{7.13} \\
&&-(\frac{\overrightarrow{\sigma }\cdot \overrightarrow{p}\phi _{o}}{2m_{o}c}%
)^{\dag }c\overrightarrow{\sigma }\cdot \overrightarrow{p}\phi _{o}+i\hbar
2m_{o}c^{2}(\frac{\overrightarrow{\sigma }\cdot \overrightarrow{p}\phi _{o}}{%
2m_{o}c})^{\dag }(\frac{\overrightarrow{\sigma }\cdot \overrightarrow{p}\phi
_{o}}{2m_{o}c})  \nonumber
\end{eqnarray}%
Since $(\frac{\overrightarrow{\sigma }\cdot \overrightarrow{p}\phi _{o}}{%
2m_{o}c})^{\dag }=\frac{1}{2m_{o}c}\phi ^{\dag }(\overrightarrow{\sigma }%
\cdot \overrightarrow{p})^{\dag }$, eq.(\ref{7.13}) reads

\begin{equation}
\widetilde{L}_{D}(\phi _{o})=i\hbar \phi _{o}^{\dag }\partial _{t}\phi _{o}-%
\frac{1}{2m_{o}}\phi _{o}^{\dag }p^{2}\phi _{o}  \label{7.14}
\end{equation}%
where we have used $(\overrightarrow{\sigma }\cdot \overrightarrow{p})(%
\overrightarrow{\sigma }\cdot \overrightarrow{p})=p^{2}$. In particular, the
last term of eq.(\ref{7.14})\ can be written as

\begin{equation}
\phi _{o}^{\dag }p^{2}\phi _{o}=\hbar ^{2}\phi _{o}^{\dag }\bigtriangledown
^{2}\phi _{o}=\hbar ^{2}((\overrightarrow{\bigtriangledown }(\phi _{o}^{\dag
}\overrightarrow{\bigtriangledown }\phi _{o})-\overrightarrow{%
\bigtriangledown }\phi _{o}^{\dag }\overrightarrow{\bigtriangledown }\phi
_{o})  \label{7.15}
\end{equation}%
The first term of the r.h.s. of eq.(\ref{7.15})\ is a divergent term, which
only contributes with a surface term that becomes zero when the Lagrangian
density is integrated. Then, eq.(\ref{7.14})\ finally results%
\begin{equation}
L_{NR}=\widetilde{L}_{D}(\phi _{o})=i\hbar \phi _{o}^{\dag }\partial
_{t}\phi _{o}-\frac{\hbar ^{2}}{2m_{o}}\overrightarrow{\bigtriangledown }%
\phi _{o}^{\dag }\overrightarrow{\bigtriangledown }\phi _{o}  \label{7.16}
\end{equation}%
which is the desired non-relativistic Schr\"{o}dinger Lagrangian. In turn,
the Euler-Lagrange non-relativistic equation reads%
\begin{equation}
i\hbar \partial _{t}\phi _{o}=\frac{\hbar ^{2}}{2m_{o}}\overrightarrow{%
\bigtriangledown }^{2}\phi _{o}  \label{7.17}
\end{equation}

\subsection{The full group of relativistic quantum mechanics with $U(1)$
fields}

With respect to the space-time symmetries, we can see that:

\begin{itemize}
\item The relativistic Lagrangian $L_{D}$ of eq.(\ref{7.1}) is invariant
under the Poincar\'{e} group $ISO(1,3)$, and also under its trivial
extension $ISO(1,3)\times \left\langle M\right\rangle $.

\item The non-relativistic Schr\"{o}dinger Lagrangian $L_{NR\text{ }}$ of
eq.(\ref{7.16}) is invariant under the Galilei group $\mathcal{G}$, and also
under its trivial extension $\mathcal{G}\times \left\langle M\right\rangle $%
, which, as we have seen, can be obtained from the In\"{o}n\"{u}-Wigner
contraction of $ISO(1,3)\times \left\langle M\right\rangle $.
\end{itemize}

With respect to the internal symmetries, in turn:

\begin{itemize}
\item The relativistic Lagrangian $L_{D}$ of eq.(\ref{7.1}) is invariant
under a global gauge-transformation $U(1)$ acting on the spinors. If we want
to preserve invariance under a local gauge-transformation, we have to
introduce gauge-fields, which turn out to be the electromagnetic potentials
obeying the Maxwell equations (see \cite{greiner}).

\item The non-relativistic Schr\"{o}dinger Lagrangian$\ L_{NR\text{ }}$ of
eq.(\ref{7.16}) is also invariant under a global gauge-transformation $U(1)$%
, and it is also invariant under a local gauge-transformation by means of
the introduction of electromagnetic fields (see \cite{Sujeeva}).
\end{itemize}

The fact that electromagnetism can be obtained from a non-relativistic
theory might sound weird. However, nowadays it is clear that the structure
of the Maxwell equations is not determined by the symmetry properties of
space-time but by the properties of gauge-symmetries. The only difference
between the non-relativistic and the relativistic cases is that, whereas the
gauge-potentials in the relativistic Lagrangian are strictly electromagnetic
potentials, the gauge-potentials in the non-relativistic Lagrangian are
components of a Galilean vector field, and this means that they transform as
irreducible representations of the central extension of the Galilei group.
As a consequence, in the non-relativistic case the potentials are the
\textquotedblleft magnetic limit\textquotedblright\ or the \textquotedblleft
electric limit\textquotedblright\ of the electromagnetic potential (see \cite%
{levy1}).

Now the question is how the kinematical Poincar\'{e} group and the internal
gauge-group combine together to lead to a new group, whose Casimir operators
would represent the actual-valued observables of the relativistic quantum
system according to our MHI. The point is relevant because the Casimir
operators of that new group might be different than the Casimir operators of
the component groups ($M$ and $S^{2}$ coming from the Poincar\'{e} group,
and $Q$ coming from the gauge group). Fortunately, this is not the case:
according to the Coleman-Mandula theorem (\cite{coleman}, for a simpler
presentation, see \cite{kaku}, \cite{weinberg1}), there is no non-trivial
union of the Poincar\'{e} group and the internal group. In other words, the
only possible combination between the two groups is the direct product.

Summing up, in this case the full group is%
\begin{equation}
ISO(1,3)\times \left\langle M\right\rangle \times U(1)  \label{7.18}
\end{equation}%
whose Casimir operators are those of the trivially extended Poincar\'{e}
group $-$given by the mass $M$ and the spin $S-$ and of the internal
gauge-group $U(1)$ $-$given by the charge $Q-$. In turn, the full group of
non-relativistic quantum mechanics is%
\begin{equation}
\mathcal{G}\times \left\langle M\right\rangle \times U(1)  \label{7.19}
\end{equation}%
whose Casimir operators are those of the central extended Galilei group $-$%
given by the mass $M$, the internal energy $W$, and the spin $S-$ and of the
internal gauge-group $U(1)$ $-$given by the charge $Q-$. The In\"{o}n\"{u}%
-Wigner contraction applied to the full group of eq.(\ref{7.18}) leads to
the full non-relativistic group:%
\begin{equation}
ISO(1,3)\times \left\langle M\right\rangle \times U(1)\rightarrow \mathcal{G}%
\times \left\langle M\right\rangle \times U(1)  \label{7.20}
\end{equation}%
As a consequence, according to our MHI, the actual-valued observables in the
relativistic case are $M$, $S^{2}$ and $Q$, with their actual values: mass $%
m_{0}$, spin $s$ and charge $e$. This is the result that one expects from a
physical viewpoint, since mass, spin and charge are properties supposed to
be always possessed by any quantum system and measurable in any physical
situation, and their values are precisely what define the different kinds of
the elemental particles of the theory.

\subsection{The many-particle case}

As it is well known, the many-particle case cannot rigorously treated by
relativistic quantum mechanics, and this fact leads us to the realm of
quantum field theory. We also know that, in general, the particle number is
not a conserved quantity in quantum field theory, nor a Casimir operator of
the relevant symmetry group of the theory. Nevertheless, in the particular
cases of the \textquotedblleft in\textquotedblright\ and the
\textquotedblleft out\textquotedblright\ stages of the scattering process,
it is assumed that the system can be modelled as a collection of $N$
non-interacting particles (the experimentally detected particles).
Therefore, at those stages the relevant group is the tensor product of $N$
copies of the full group $ISO(1,3)\times \left\langle M\right\rangle \times
U(1)$ (for simplicity we will only consider collections consisting of a
single kind of elementary particles). Since the representations of this $N$%
-tensor product can be expressed as products of the representations of the
factor groups, they are labelled by $N$ and the Casimir operators of the
group $ISO(1,3)\times \left\langle M\right\rangle \times U(1)$. This means
that, in the \textquotedblleft in\textquotedblright\ and the
\textquotedblleft out\textquotedblright\ stages, the particle number
operator $N$ becomes an extra Casimir operator to be taken into account.

\smallskip Let us consider a particle labelled by the Casimir operators $%
C_{1}^{PE}=M$, $C_{2}^{PE}=M^{2}$ and $C_{4}^{PE}=S^{2}$ of the trivially
extended Poincar\'{e} group $-$in the reference frame of the center of mass$%
- $ and the Casimir operator $C_{1}^{U}=Q$ \ of the gauge-group. A
non-interacting $n$-particle state is given by

\begin{equation}
|n\rangle =|1\rangle \otimes |1\rangle \otimes ...|1\rangle =(a^{\dagger
}\otimes a^{\dagger }\otimes ...a^{\dagger })|0\rangle =(a^{\dagger
})^{n}|0\rangle ,  \label{7.21}
\end{equation}%
where $a^{\dagger }$ is the creation operator of each particle, and 
\[
N|n\rangle =n|n\rangle , 
\]%
is the particle number, which can be easily seen to be additive. Now we can
combine the already known Casimir operators, and define the new operators
for the collection of $N$ particles with the same mass and spin: 
\begin{equation}
\text{Mass=}(C_{2}^{PE})^{\frac{1}{2}}N\text{ \ \ \ Spin=}\left(
C_{4}^{PE}\right) ^{\frac{1}{2}}\text{ \ \ \ Charge=}Q\text{\ \ \ \ Particle
Number=}N.  \label{7.22}
\end{equation}%
According to our MHI, all these operators represent actual-valued
observables of the system of $N$ non--interacting particles. Let us stress
again that this result is not general in quantum field theory, but it is
only valid in the \textquotedblleft in\textquotedblright\ and
\textquotedblleft out\textquotedblright\ stages of the scattering process.
Nevertheless, the fact that $N$ turns out to be a definite-valued observable
is a reasonable result in those stages, where the number of particles is
always considered a definite magnitude of the system. In turn, since in the
interacting stage $N$ is not a Casimir operator, according to our MHI it is
not a definite-valued observable; and this is also reasonable because the
particle number is not expected to be definite in the presence of
interaction.

\section{Conclusions}

The interpretation of quantum mechanics is still one of the most debated
problems in the foundations of physics. Although many new formal results
were obtained during the last decades, the links with physical models have
lost their strength in the discussions. With our MHI, we have tried to
revert this situation by taking into account physical observables and their
physical meaning as generators of symmetries. In particular, we have
expressed our rule of actual-value ascription in a Galilei-invariant form,
in terms of the Casimir operators of the Galilei group.

On the other hand, the interpretation of quantum relativistic theories has
been a much less discussed topic. In this paper we have argued that, since
group considerations play a central role in our interpretation of
non-relativistic quantum mechanics from the very beginning, the
extrapolation of the strategy to the relativistic case $-$by replacing the
relevant group$-$ is straightforward: the actual-valued observables of the
system turn out to be the Casimir operators of the Poincar\'{e} group and of
the internal gauge-group. In particular, we have shown that this
extrapolation leads to physically reasonable results, since the
actual-valued observables so selected are magnitudes supposed to be always
possessed by the systems, and they are also the properties that contribute
to the classification of elemental particles. Moreover, we have also proved
that the actual-valued observables in the relativistic and the
non-relativistic theories are correctly related through an adequate limit,
which can also be expressed in group terms. On the basis of these results,
we consider that the further extension of the MHI to quantum field theory is
an issue that deserves to be studied.

\section{Acknowledgments}

We are particularly grateful to Rutwig Campoamor-Stursberg for stimulating
discussions. This work was partially supported by grants of the Buenos Aires
University, the CONICET and the FONCYT of Argentina.

\end{document}